\documentclass[journal]{IEEEtran}
\usepackage{graphicx}
\usepackage{cite}
\usepackage{float}

\begin{document}

\title{A Mobile Food Recommendation System Based on The Traffic Light Diet}
\author{
\IEEEauthorblockA{Thienne Johnson\textsuperscript{1,2}, Jorge Vergara\textsuperscript{2} , Chelsea Doll\textsuperscript{3}, Madison Kramer\textsuperscript{3}, Gayathri Sundararaman\textsuperscript{2}, \\  Harsha Rajendran\textsuperscript{2}, Alon Efrat\textsuperscript{2} and Melanie Hingle\textsuperscript{3}}

\IEEEauthorblockA{ \textsuperscript{1} Dept. of Electrical and Computer Engineering, \textsuperscript{2} Dept. of Computer Science, \\
\textsuperscript{3} Dept. of Nutritional Science\\
University of Arizona, Tucson, AZ 85710, USA\\
\{thienne,jbv, gayathrisram, hvrajendran, alon, cdoll, mkramer, hinglem\}@email.arizona.edu\\}
\thanks{CREU (Collaborative Research Experience for Undergraduates) final report, May 2013. http://foodtracker.cs.arizona.edu}}

\maketitle

\begin{abstract}
Innovative, real-time solutions are needed to address the mismatch between the demand for and supply of critical information to inform and motivate diet and health-related behavior change.  Research suggests that interventions using mobile health technologies hold great promise for influencing knowledge, attitudes, and behaviors related to energy balance. The objective of this paper is to present insights related to the development and testing of a mobile food recommendation system targeting fast food restaurants. The system is designed to provide consumers with information about energy density of food options combined with tips for healthier choices when dining out, accessible through a mobile phone.

%\keywords{mobile, parser, location, energy density, dietary behavior, health, recommendation}
\end{abstract}

\section{Introduction}

Innovative, real-time solutions are needed to address the mismatch between the demand for and supply of critical information to inform diet and health-related behavior change.

Widespread use of the Internet has provided an opportunity to expand the reach of health education programs, as well as provide continuous support and tools for achieving necessary changes in multiple lifestyle behaviors, e.g. healthful eating, regular physical activity, and management of medications \cite{1}. The popularity of mobile platforms (including cell-phone based applications which allow direct access to the Internet) and the increasing use of social networks allow broad adoption of a new range of applications and services focused on promoting mobile health (mHealth) and well-being \cite{2}. mHealth platforms are being developed to address health concerns such as diabetes, food allergies, and high blood pressure. As mHealth interventions also offer some level of anonymity to users, this may encourage individuals to seek out sensitive health information and “ask” questions that they wouldn't normally ask or feel comfortable asking their healthcare provider \cite{3}.

While health applications using mobile platforms have become a reality, data related to whether and how individuals access specific health information, and importantly, how they use information to make health-related decisions remains almost entirely within the private sector. The little empirical data that are available suggest lower user acceptance \cite{4} than expected given the current position of the mobile apps market in the hype cycle. Given the potential for broad population reach of mobile interventions, it is critical to gain a better understanding of the ways in which consumers wish to access health information, and how they use this information to make changes in health-related behaviors such as the purchase and consumption of food.

There are is a significant number of mobile applications (apps) available for smartphones targeting diet. The majority of applications are designed to help the consumer count calories (e.g., Calorie Count, Diet Assistant, My Diet Coach, among several others.), track their intake (Diet Diary, Doc’s Diet Diary, etc.) or follow a specific plan (hCG diet, Atkins Diet Shopping List, the 90 Day Diet, etc.). However, limited research has tested whether or not the use of these applications elicits changes in dietary behavior, and which aspects of the applications (i.e., calorie counting, tracking, or following a specific plan) – if any - are most effective.

Research suggests that changing dietary behavior does not involve ``unlearning'' old habits, but rather, learning new ones, and that the context in which learning takes place (both physical and social) significantly influences the learning process \cite{5}. Thus applications designed to help individuals engage in specific dietary behaviors should acknowledge and address this context, providing critical information and feedback when and where eating behaviors are occurring.

The purpose of our project is to develop a food recommendation system that would allow researchers and students from multiple disciplines (Computer Science, Nutritional Sciences, Public Health) to study and understand common challenges related to the function and user acceptance of mobile health applications designed to help consumers change dietary behavior. Research topics include issues surrounding data collection and privacy, data representation, the user interface, design patterns, mapping between current location and a food intake place, content accuracy, and perceived relevance of the information to the location/situation.

The contribution of our findings include new insights on mobile computing applications for health behavior change, such as the use of location-awareness combined with a recommendation database to assist users in making informed choices when eating out. We also propose the use of ``easy to understand and associate'' icons and colors from the Traffic Light Diet system in the mobile environment, making it very easy for users of any age to understand if a food item is ``good'' or ``bad''. We also make the results from user studies available so mobile systems developers targeting diet applications can use the information as an initial guide.

The rest of the paper is organized as follows: Section \ref{sec:architecture} details the construction of the system architecture, composed of servers, database, the guiding framework for nutrition recommendations and the mobile app. Section \ref{sec:evaluation} describes user testing and evaluation of the system. Section \ref{sec:discussion} discuss some options for improving food recommendation system for mobile users.  Finally, section \ref{sec:conclusions} presents conclusions and future work.

\section{System Architecture}\label{sec:architecture}

This project relied on mobile users accessing restaurant menus, food items nutritional information and tips for healthier eating. The system architecture is composed by a database containing all the information to be displayed, a mobile user application (the ''Foodtracker'' app), web interface for nutritionists and administrators.

This section presents technical details about the system architecture and choices made to improve system’s availability and security.

\subsection{Web server and services}\label{sec:servers}

A Foodtracker user may request restaurant menus at any moment of the day, at any location. The choice of creating an application with a large database to include all possible restaurants would be a poor one, because the device would have to store (most of the time) unnecessary information. Also menus are subject to change (inclusion or removal of food items) and need to be updated periodically.

\begin{figure}
\centering
\includegraphics[height=4.5cm]{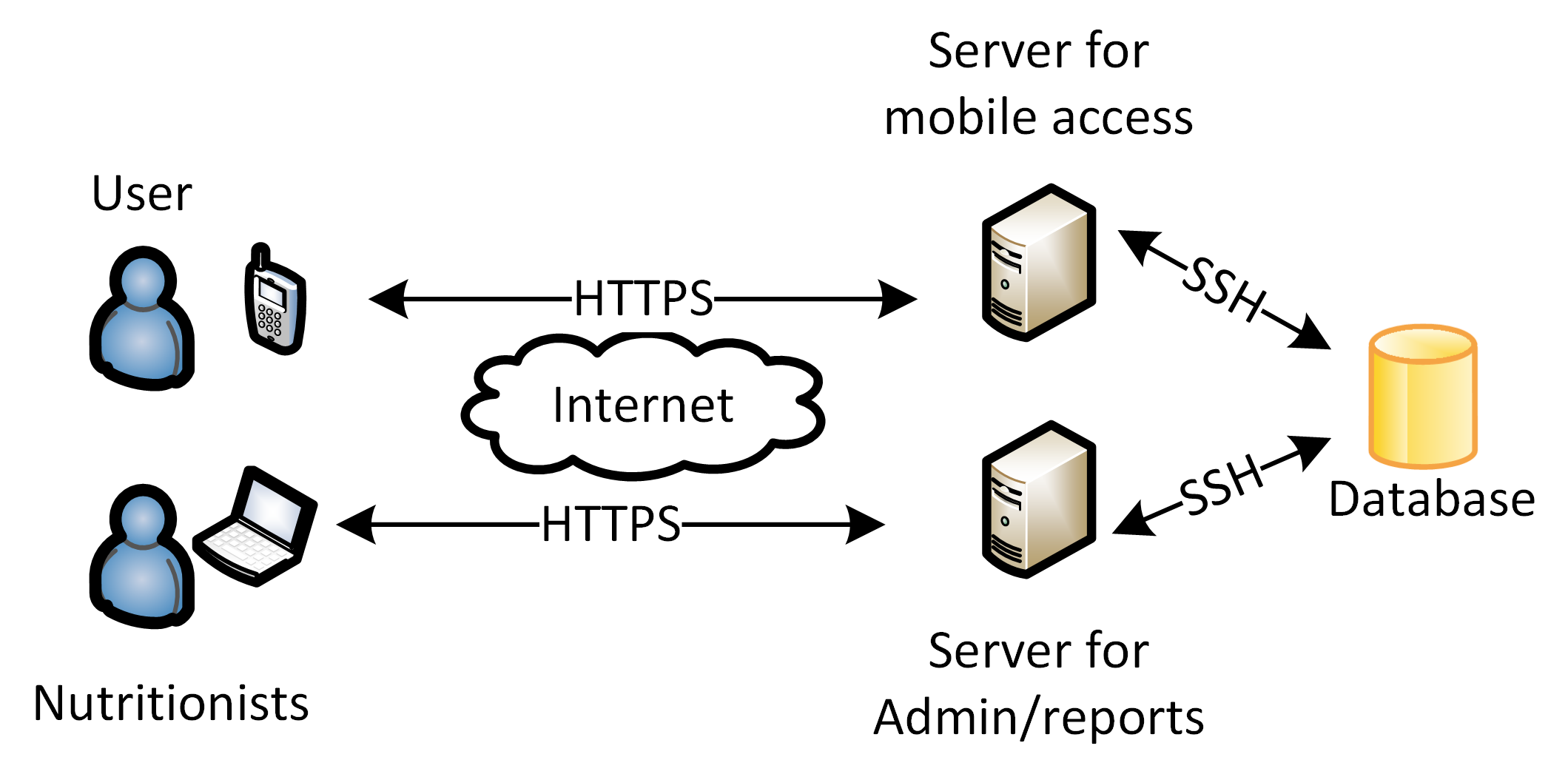}
\caption{System architecture}
\label{fig:system}
\end{figure}

A food database was created and contains information on restaurants, food categories (Burgers, Pizza, Salad, Desserts, etc.), and food items and their nutritional information. This database can be accessed from the mobile app but also is accessible to nutritionists and system administrators. Thus, two separate Tomcat Apache web servers were created to handle requests from different types of users and each Tomcat instance has its own set of Java Servlets. This separation was designed to increase system availability in case of failure of one of them servers. Each Tomcat server communicates with a MySQL database instance that stores restaurant menus, admin users and is ready to store mobile user information (see section \ref{sec:conclusions}) (Figure \ref{fig:system}).

The communication with the servers is performed over HTTPS (Hypertext Transfer Protocol Secure) connections and the connection to the Database is performed on SSH (Secure Shell Protocol) sessions.

\begin{figure}
\centering
\includegraphics[height=3.5cm]{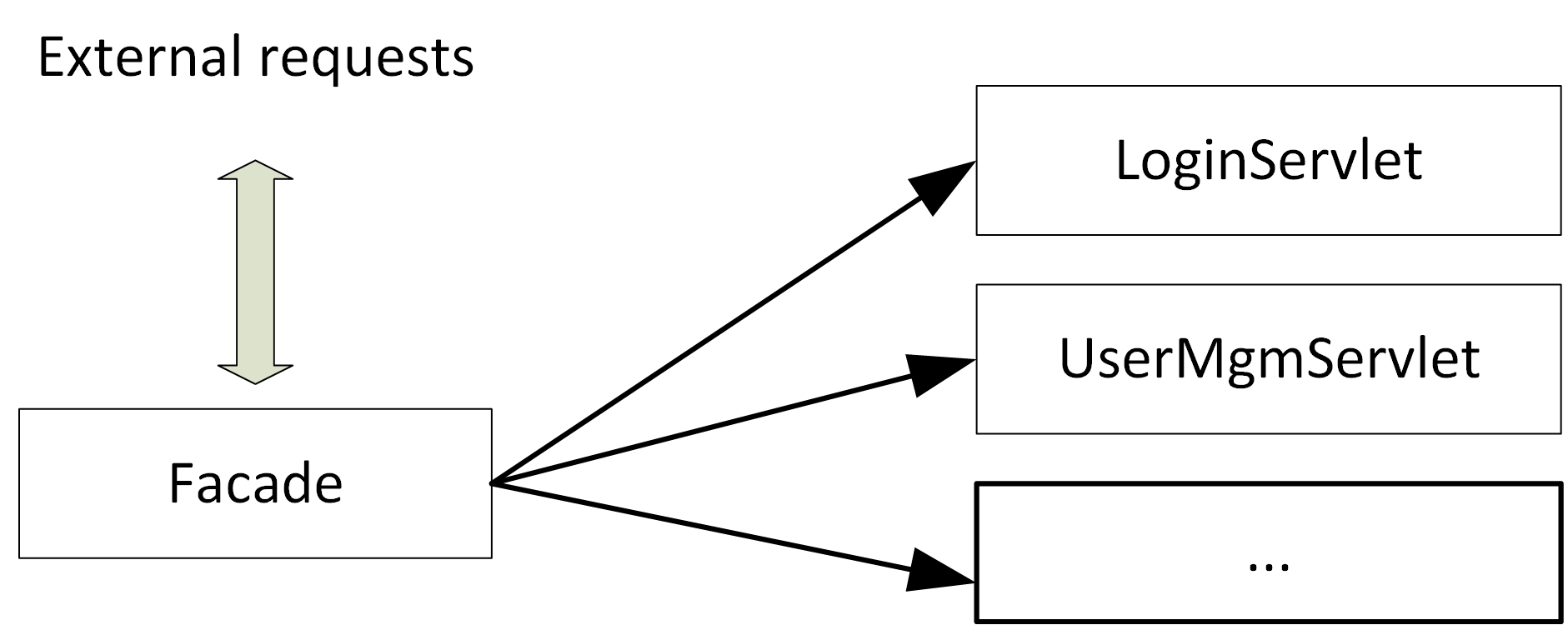}
\caption{Facade for servlets}
\label{fig:facade}
\end{figure}

The Facade design pattern \cite{9} was used in the admin web servlets in order to provide a unified interface to a set of interfaces in a subsystem. Facade defines a higher-level interface that makes the subsystem easier to use and also do not expose the name of the internal servlets in the JSP pages handled by the Tomcat Servlets. Each call to the internal subsystem was executed after an Authorization/Authentication phase (Figure \ref{fig:facade}), thus non-authorized users can not use the system. The servlets handling the requests from the mobile users do not need this type of session handling because they are not sending or acessing sensitive information.

\subsection{Automated Menu Parser}\label{sec:parser}

Fast food restaurants menus are available online in several restaurants and diet related websites. Instead of entering data manually into the food database, we chose the approach of building an automated parser in order to populate the database with fast food restaurant menus.

As the web is a vast growing source of information, which most of it is in the form of unstructured text, the available information hard to query \cite{9}. Online pages created to display nutritional information for restaurants do not employ a structured format to display the data. As the construction of an automated parser to acquire the target information needed inspection of how data was displayed with HTML tags (“$<$tagname$>$”), the rules for parsing automatically could be explored. The parser resides in the administrator server, and a system administrator can execute the parser servlets. The servlets uses the JSoup API \cite{jsoup}, which is a Java library created to extract and manipulate data, using DOM, CSS, and JQuery-like methods.

\begin{figure*}[ht]
\centering
\includegraphics[height=10cm]{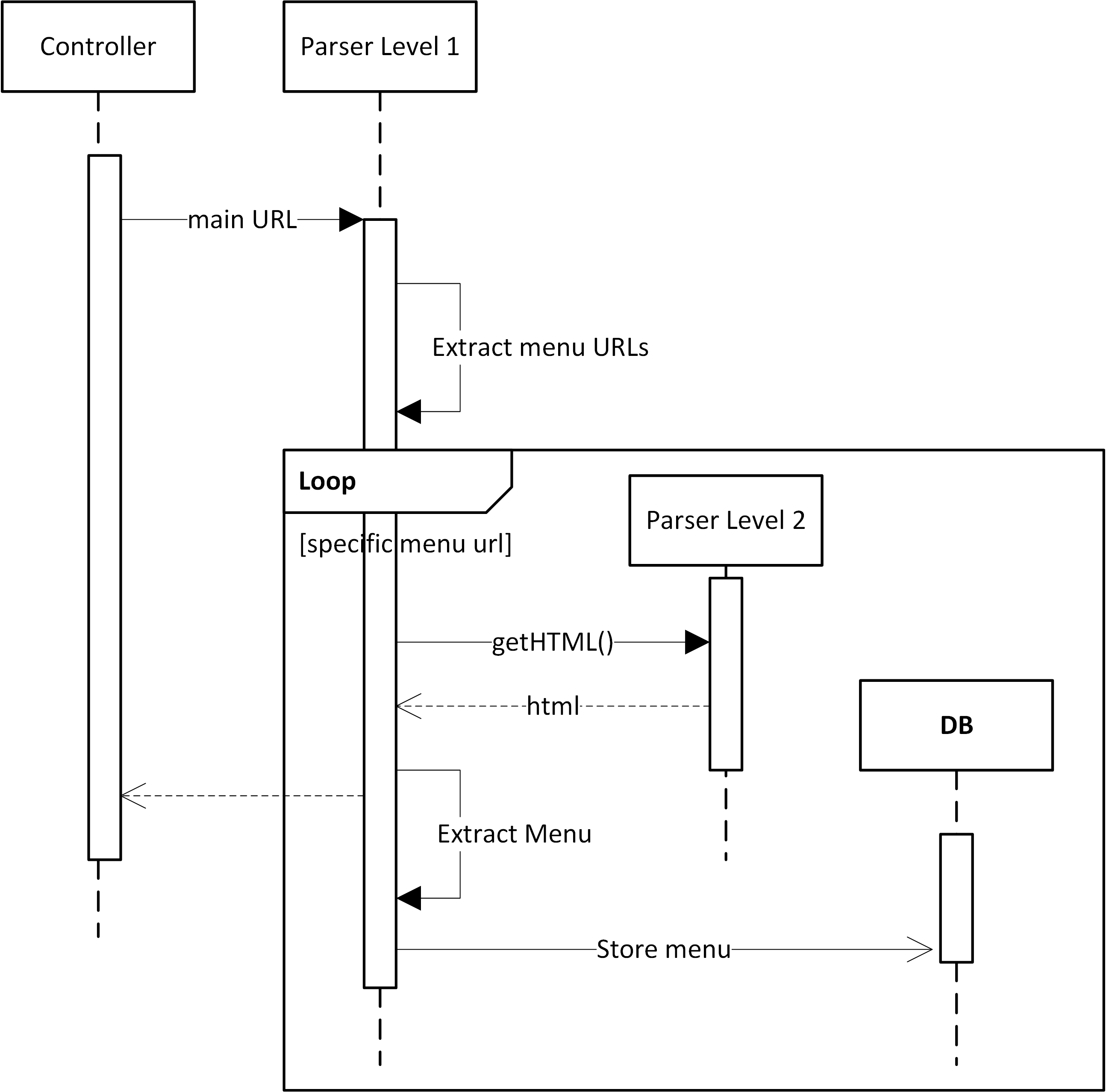}
\caption{Automated menu parser sequence diagram}
\label{fig:parser}
\end{figure*}

The html parsing is a three steps process, depicted in the sequence diagram presented in Figure \ref{fig:parser}. A Controller class is responsible to get from the administrator URLs (Uniform Resource Locator) that identifies the root (or main page) of menu repositories.

For the current implementation we worked with one website  containing the needed data\footnote{http://www.exercise4weightloss.com/weight-watchers-points.html}. The website includes nutrition information for many restaurants, including calories, total fat, saturated fat, dietary fiber, protein, carbohydrates and sodium. Not all the restaurants in the data source website contain the same information (some restaurants offer less nutritional info fields), so the parser was created to be adaptable to such situations.

After receiving the URL, the Controller calls the \textit{Parser Level 1} class, which is responsible for extracting the data from the URL, and needs to find the URLs that points to each restaurant menu. For each new URL found, a \textit{Parser Level 2} object is created. It acquires the HTML information, and looks for specific columns orders to extract a) how the nutritional information is organized and b) the nutritional value for each item.

\begin{figure*}
\centering
\includegraphics[height=9cm]{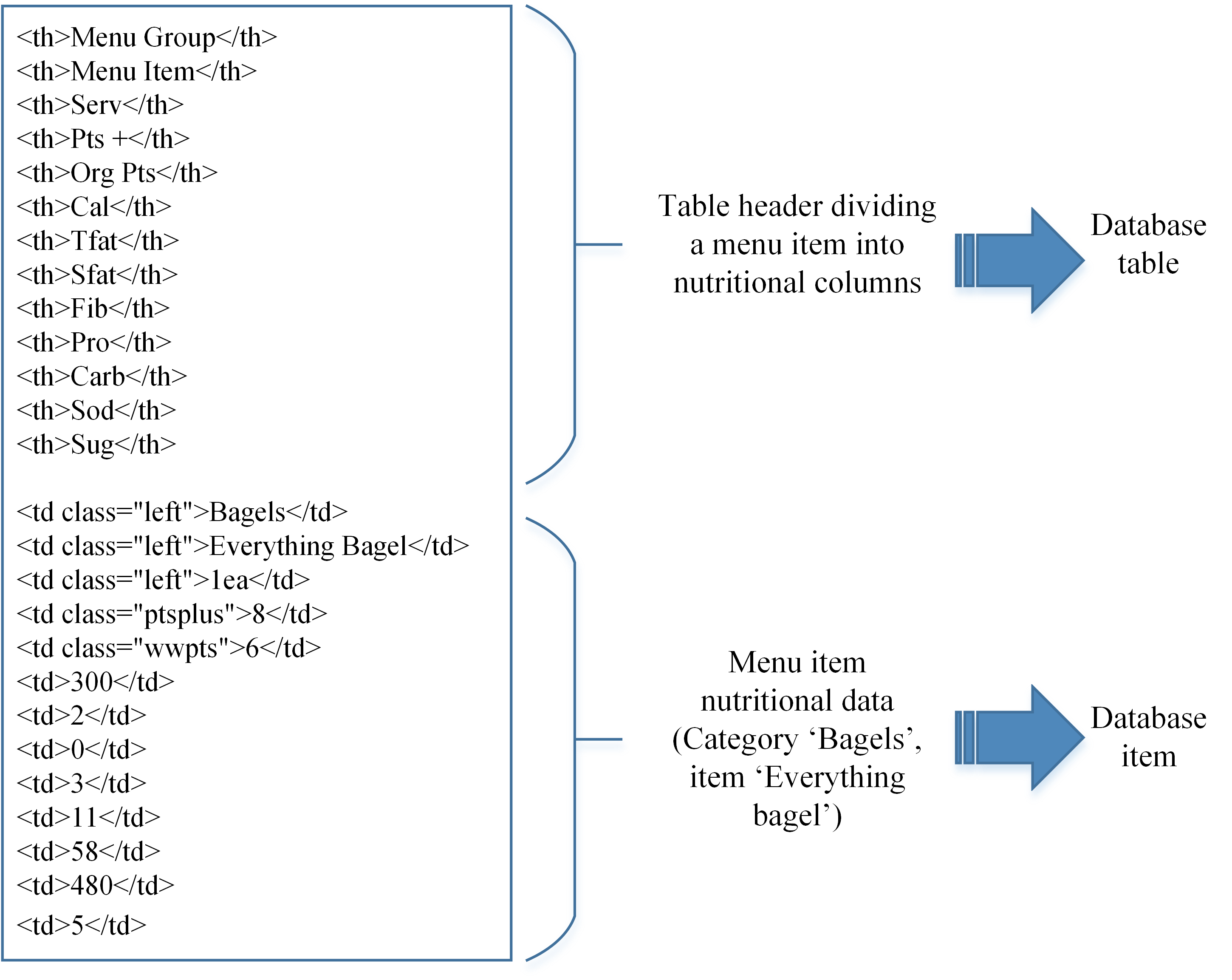}
\caption{HTML table elements}
\label{fig:htmltable}
\end{figure*}

Figure \ref{fig:htmltable} summarizes one example of items a) and b). \textit{Parser Level 2} needed to be adjustable to table formats. HTML restaurant tables for each restaurant may not follow the same order when displaying nutritional items. When extracting each menu item and storing the information in the database, this order needs to be verified. After extracting a menu, the information is saved in the database. In this example, lines 1-13 containing the columns headers (represented by the tag $<th>$), thus this information is used to track which columns are of interested for further storage. Lines 14-26 contain the table's contents for each column (represented by the tag $<td>$). Thus the food category is ``Bagels'', item ``Everything bagel'', and its nutritional information in the remaining columns.

After finishing parsing the restaurant menu, the system returns to \textit{Parser Level 1}. If there are URLs in its queue, it sends a new URL to be extracted by \textit{Parser Level 2}, until it reaches the last restaurant URL.

\subsection{Foodtracker App}\label{sec:app}

A mobile application called ``Foodtracker'' was developed for the Android platform. The choice of opting for native development was due to the need to access to the GPS sensor on the device.

The app uses the Traffic Light Diet (detailed next) as basis for the food intake recommendation system. The use of graphical signs help the facilitate the association of colors/signs to good (healthy) or bad (unhealthy) food quality.

\subsubsection{The Traffic Light Diet}\label{sec:diet}

Aspects of Epstein’s Traffic Light Diet (TLD) \cite{9} were adapted to categorize fast food selections based on energy density. TLD broadly categorizes foods on the basis of their macronutrient content into the colors of a traffic light with similar meaning:  “Green” foods are low in calories and fat ($<2$g of fat per serving); “Yellow” foods are medium-calorie and fat foods (between 2 and 5g of fat per serving); and “Red foods” are high in calories and fat ($>5$g of fat per serving).

The TLD categories were chosen because they are common and widespread symbols, which intuitively translate to nutrition guidelines that suggest “eat more” (Green) “eat moderately” (Yellow) and “eat less” (Red), without additional explanations or prior nutrition knowledge.

Table 1 summarizes the three categories, presenting each of the three symbols used in the mobile app and the parameters used to classify food items into the categories. Each food item will receive a symbol based on the fat content when displaying the restaurant menu.

\begin{table}[ht]
\caption{Labeling system in the App}
\centering
\begin{tabular}{c c} % centered columns (2 columns)
\hline %inserts double horizontal lines
Symbol          & Fat content per serving \\
\includegraphics[height=1cm]{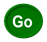}  & Less than 2g \\
\includegraphics[height=1cm]{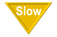} & Between 2 and 5g\\
\includegraphics[height=1cm]{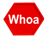}     & Greater than 5g  \\% [1ex] adds vertical
\hline %inserts single line
\end{tabular}
\\
\label{tab:labels}
\end{table}

\subsubsection{App Details}\label{sec:appdetails}

\begin{figure*}
\centering
\begin{minipage}{.5\textwidth}
  \centering
\includegraphics[height=8cm]{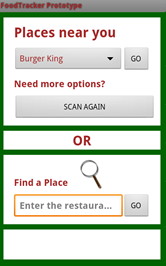}
\end{minipage}%
\begin{minipage}{.5\textwidth}
  \centering
\includegraphics[height=8cm]{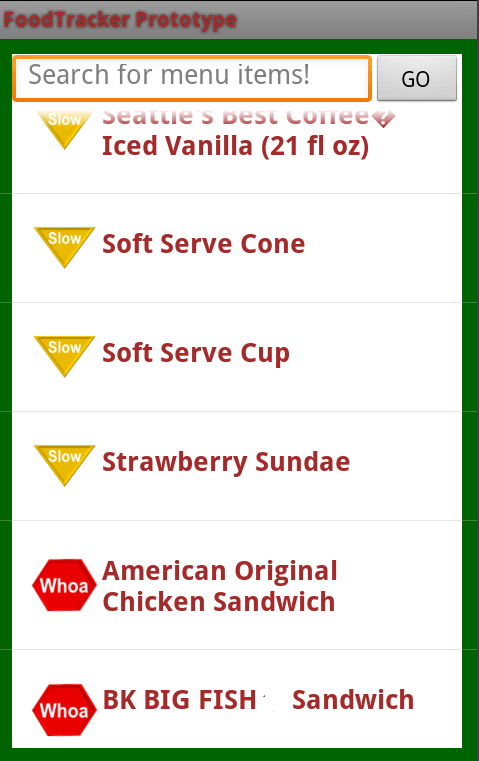}
\end{minipage}
\label{fig:app}
\caption{Screenshots from the app}
\end{figure*}

The Foodtracker presents restaurant menus and their nutritional information. When the user starts the app, he/she can choose to see menus by categories (Burger, Salad, etc) or by location (Figure \ref{fig:app}a). The ``Place near you'' box offers a drop down element with restaurants located less than 500 meters from the user. Google Maps API \cite{gmaps} was used to track the restaurants closest to the current user location.  This option is easily reconfigurable in the API so the range can be adjustable. The option ``Scan again'' will present an updated list for the current location. The user can also find a restaurant menu by giving the restaurant name in the bottom box.

For each restaurant, the menus are presented by ordering the food items. First, green items are shown, then yellow and red (Figure \ref{fig:app}b). We chose this approach to present ``better'' food items first, so the user understand that, by going down the list he/she is looking into less healthier items. By clicking on any item, the nutritional content for that item is displayed. In the current system, we are displaying Calories, Fat and Saturated Fat. Other nutritional information could be displayed (because the parser extracts and stores all the available nutritional information for each menu it extracts), but for the current study we decided to use only these parameters.

\begin{figure*}[ht]
\centering
\includegraphics[height=8cm]{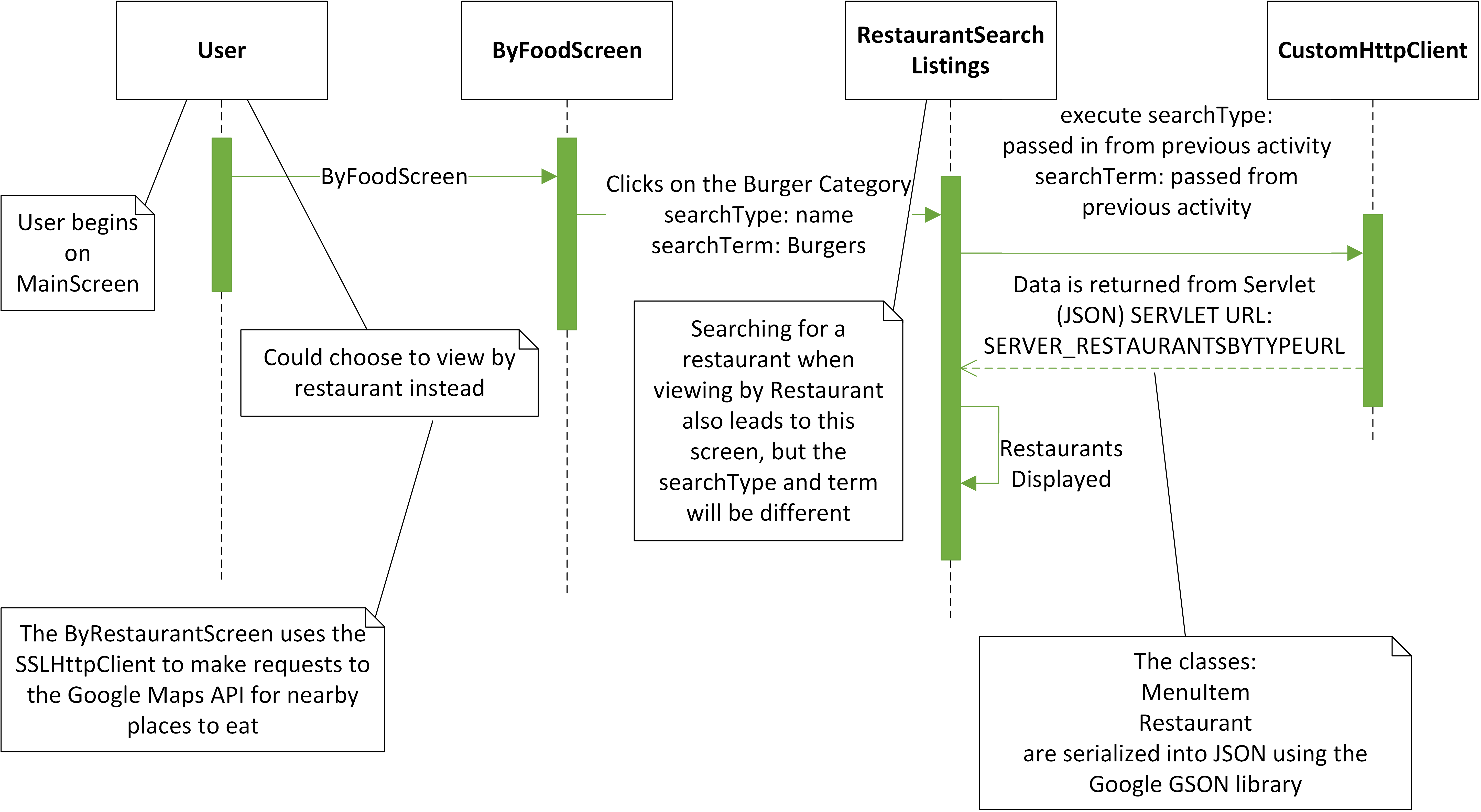}
\caption{ Sequence diagram (1) to present information to the user}
\label{fig:sequence1}
\end{figure*}

Figure \ref{fig:sequence1} presents a sequence diagram representing the app execution when the user requests to filter content based on categories. All classes in this diagram reside in the mobile system.

The user selects ``By food type'' screen, which offers food categories (Burgers, Sandwiches, etc), instead of restaurant names. After choosing category ``Burger'', the request is sent to the server, which will return a JSON (JavaScript Object Notation, a standard for data exchange) array (using the Google-GSON library\cite{gson}) with the names of the restaurants who offer ``Burgers''. All the network requests in the app are handled by the \textit{CustomHTTP} class for a cleaner remote network access code in the app.  The menu for the selected restaurant is retrieved and presented.

\begin{figure*}[ht]
\centering
\includegraphics[height=7.5cm]{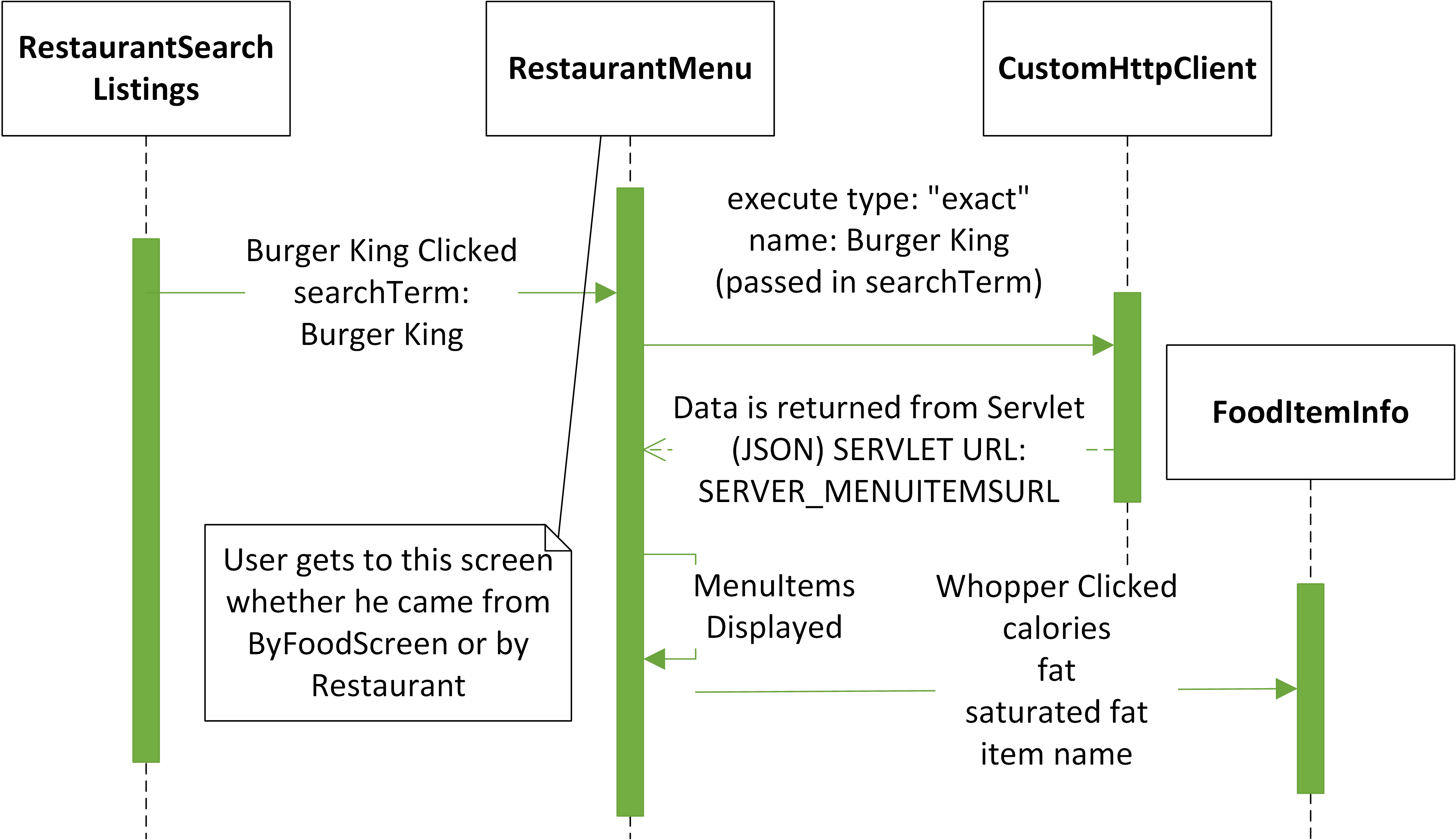}
\caption{ Sequence diagram (2) to present information to the user}
\label{fig:sequence2}
\end{figure*}

Figure \ref{fig:sequence2} represents the sequence after the previous retrieval. For the presented restaurant list, the user chose ``Burger King'', and a new request is sent to the server. The server returns a new JSON array containing the menu items for this restaurant.

\section{Evaluation}\label{sec:evaluation}

\subsection{Pre testing phase}
During the design and implementation stage of the system, Nutritional Sciences students developed a questionnaire designed to ``pre-test'' the concepts underlying the recommendation system. Pre-test questions were developed to elicit information about consumers’ notions about making healthy choices in fast food restaurants, and whether and what type of nutrition information or labeling might help consumers select specific foods.

Ten volunteers answered questions related to healthy eating at fast food restaurants and nutrition labeling, and findings were used to refine the app’s structure, features, and function and to inform the procedures and questions used during the beta testing phase (Table  \ref{tab:pretest}).

\begin{table}
\caption{Pre-testing Questions}
\centering
\fbox{
\parbox{9cm}{
\textbf{What health related applications, if any, do you currently use on your phone? (Which apps do you like already and why they are currently using them. Helps with a model of our application.)}
\begin{itemize}
\item  calorie counter, Nike plus
\item Don’t use any. If I were to think of an ideal app it would include calorie counter and information that’d help me make healthy decisions throughout the day
\item None, would want calorie counter, maybe one that has a healthy restaurant locator
\item None, would want an exercise tracker
\item Points on weight watchers (hard to keep up with)
\item I don't even have a smart phone!
\item I don't have one. Though I would like a calorie counter or an app where I can put the ingredients I have and it gives me a healthy recipe.
\end{itemize}

\textbf{ Think about one of the health applications that you use often, what do you like about them? (usability, useful info, features that make this app popular)}
\begin{itemize}
\item  tracks daily caloric intake
\item  Like that it gives you your total allowance at beginning of day, what you can and can't have. If eat bad know that you only have a certain amount of points left
\end{itemize}

\textbf{Do you think healthy decisions can be made at fast food restaurants? What would make it easier?}
\begin{itemize}
\item Menu labeling that clearly identifies healthier choices
\item Separate menus for lower calories items
\item Calorie counts and other relevant nutrition information next to each of the items
\item Decrease portion sizes
\item Calorie counts help, but just because something is low in calories doesn't mean that it’s healthy
\end{itemize}
\textbf{How does reading a nutrition label help you to understand the food choices you’re making? How can reading these be made easier? }
\begin{itemize}
\item Companies should make labels easily understood for non-experts (maybe have additional information on a website)
\item Providing information on daily amounts people should eat would be helpful
\item Visuals could help make it easy, e.g., like a ``calorie bar''
\item Presenting the information differently might make it easier to understand
\end{itemize}
}}
\label{tab:pretest}
\end{table}

\subsection{Beta-Testing Phase}

The application was beta-tested with ten users following an established protocol. Each user was asked to 1) open the application, 2) choose a restaurant based on their location, and 3) choose a menu item from the list. Two trained interviewers (Nutritional Sciences undergraduate research assistants) followed the test with a series of questions adapted from mHiMSS guidelines on evaluation of the usability of medical applications \cite{7} (Table \ref{tab:betatest}).

Immediately following the beta test, face-to-face interviews were conducted with ten individuals using a semi-structured script. The script was designed to include open-ended questions and data were analyzed per standard qualitative data methods. Table \ref{tab:betatest} contains questions and a summary of the related answers.

Following pre-testing, feedback from the ten volunteers was used to make final adjustments to the application. These adjustments were minor, and primarily related to size and location of specific features and graphics, e.g., location of traffic light symbol in relation to the menu item (to the left or right of the text), size of the traffic light symbol, number of clicks to obtain relevant information about each menu item, and location and accessibility of additional information (i.e. ``Tips on Making a Whoa Food into a Go Food'').

\begin{table}[ht]
\caption{Beta-testing questions}
\centering
\fbox{
\parbox{9cm}{

\textbf{What did you like best about this application?}

Colors, design simplicity, separate buttons for nutritional facts and tips; healthy foods were listed first; could search by location or restaurant; specific number of calories for people that like to track their calories; how it helps you decide how to make things healthier; the go, slow and whoa ratings; that you can find ways to eat healthier at fast food restaurants.
\par\vspace{\baselineskip}
\textbf{What didn't you like about this application?}
No sodium content available; unnecessary menu items (often a part of ‘sides’ category).
\par\vspace{\baselineskip}
\textbf{Is there something we should add to make this application better? }
Pull up restaurants besides just fast food restaurants; list carbs and other nutrients; add ingredients; add allergy warnings; make it so you can change the mile radius.
\par\vspace{\baselineskip}
\textbf{How would you use this application in everyday life, if at all? Do you see yourself using this application to help make food intake decisions, why or why not?}
Would use it when going out to eat; to pick a healthy restaurant; when I want to make healthy decisions at a fast food restaurant or when eating out; helpful when trying to find healthy options while you’re out instead of googling them; when eating a fast food place to rate the different foods.
\par\vspace{\baselineskip}
\textbf{Were the colors helpful in making your food decisions? Were the words that were used helpful in making food decisions? What would make these features easier to understand?}
Overall consensus: Yes, very helpful and easier to understand.
\par\vspace{\baselineskip}
\textbf{Would you recommend this application to friends or family, why or why not? }
Use it and recommend; wish I could get this app right now; yes, but this is a very basic app, and could use some “spicing up” before going public; it’s different than other food applications already out there; you can use it for everyday life not just people with diabetes or other health risks
}
}
\label{tab:betatest}
\end{table}

\section{Discussion}\label{sec:discussion}

\begin{figure*}[ht]
\centering
\includegraphics[height=8cm]{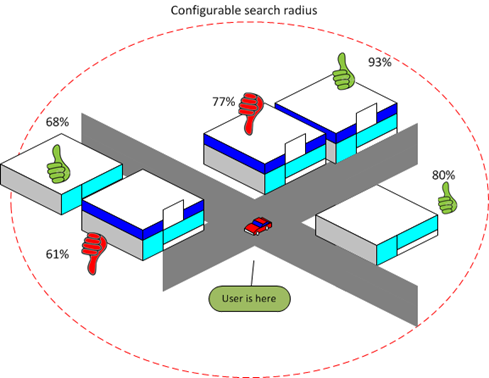}
\caption{Location context for recommendation process}
\label{fig:recommendation}
\end{figure*}

Research suggests that changing dietary behavior does not involve “unlearning” old habits, but rather, learning new ones, and that the context in which learning takes place (both physical and social) significantly influences the learning process \cite{5}. So applications designed to help individuals engage in specific dietary behaviors should acknowledge and address the learning context, providing critical information and feedback when and where eating behaviors are occurring.

The proposed system presents choices to the users. But, for now, the system is not aware of which option the user chooses and therefore it is not possible to verify if the proposed system is effective at motivating users to choose the more healthy options. If the user's choice is recorded and verified against the options presented to the user or user's historical choices, it is then possible to analyze if/how the user is changing his behavior. However, there are privacy issues if such information is recorded. Thus user needs to be aware of the tracking by explicit requests. The same issue is valid for recording user's location over time or during the app usage.

Another topic of interest is the use of ratings, such as Yelp, Urbanspoon,etc. Such services offer recommendations based on user experience (food quality, overall restaurant evaluation, etc). It'd be easy to incorporate ratings regarding food ranking (such as proposed by our project). By acquiring user's evaluation per restaurant, it is possible to give recommendations based on overall users' evaluation. To enable this, ratings could be displayed for each restaurant in the vicinity, highlighting restaurants that offer relatively more healthy options. Figure \ref{fig:recommendation} presents one option for the recommendation comparison. Restaurants closer to the user could be colored (in the app listing) with colors related to a ranking of the evaluations. In the example, green thumbs represent a good choice of healthy food items and the red color represent a restaurant with not so many good choices.

Another topic of research is the adaptation of suggestions to the specific users' profiles, for example, users with intolerance to gluten, milk, etc. In this case, each food item needs to be classified \textit{per health condition} before offering the item evaluation as good or bad.

\section{Conclusions and Future Work}\label{sec:conclusions}

The proposed system is still in its initial testing and evaluation phases, but preliminary data suggest it can be used (and is useful) for members of the community. The prototype was functional, and users’ evaluation suggested that the product is useful.

The application and database were built to be HIPAA (Health Insurance Portability and Accountability Act \cite{10}) compliant and designed to accommodate future upgrades; for example, expanding the tracking feature to allow users to document their choices and recording how many ``red'' food items they ate that day/week/month.  Related to this, visualization of food choices over time using formats that would be compelling to potential consumers of such information.

In addition, tracking choices to location could provide valuable insight into the types of foods and energy density of foods consumed at different types of fast food restaurants. Additional information to guide users could include an option for written recommendations for the current location (restaurant) and let other users read (and give their) the recommendations about the menu items or how easy/difficult it was to make healthy choices at a particular location.

\subsubsection*{Acknowledgments.}
We would like to thanks the CRA-W (CRA Committee on the Status of Women in Computing Research) and CDC (Coalition to Diversify Computing) for the CREU (Collaborative Research Experience for Undergraduates) support.

\end{document}